\input harvmac
\input amssym.def
\input amssym.tex

\noblackbox

\def\BR{{\Bbb R}}
\def\BZ{{\Bbb Z}}
\def\CH{{\cal H}}
\def\CK{{\cal K}}
\def\CL{{\cal L}}
\def\CM{{\cal M}}
\def\CN{{\cal N}}
\def\CO{{\cal O}}

\def\^{{\wedge}}
\def\*{{\star}}
\def\vol{{\hbox{vol}}}
\def\im{{\hbox{Im}}}

\def\urlfont{\hyphenpenalty=10000 \hyphenchar\tentt='057 \tt}

\newbox\tmpbox\setbox\tmpbox\hbox{\abstractfont PUPT-2025}
\Title{\vbox{\baselineskip12pt\hbox{\hss hep-th/0203061}
\hbox{PUPT-2025}}}
{\vbox{
\centerline{A Note on Fluxes and Superpotentials}
\smallskip
\centerline{In $\CM$-theory Compactifications}
\smallskip
\centerline{On Manifolds of $G_2$ Holonomy}}}
\smallskip
\centerline{Chris Beasley}
\smallskip
\centerline{\it{Joseph Henry Laboratories, Princeton University}}
\centerline{\it{Princeton, New Jersey 08544}}
\medskip
\centerline{and}
\medskip
\centerline{Edward Witten}
\smallskip
\centerline{\it{School of Natural Sciences, Institute for Advanced Studies}}
\centerline{\it{Princeton, New Jersey 08540}}
\bigskip\bigskip

We consider the breaking of $\CN=1$ supersymmetry by non-zero $G$-flux
when $\CM$-theory is compactified on a smooth manifold $X$ of $G_2$
holonomy.  Gukov has proposed a superpotential $W$ to describe this
breaking in the low-energy effective theory.  We check this
proposal by comparing the bosonic potential implied by
$W$ with the corresponding potential deduced from the eleven-dimensional
supergravity action.  One interesting aspect of this check is that,
though $W$ depends explicitly only on $G$-flux supported on $X$, $W$
also describes the breaking of supersymmetry by $G$-flux transverse to $X$.

\Date{March 2002}

\lref\achsp{B.~S. Acharya and B. Spence, ``Flux, Supersymmetry, and M-theory 
on 7-Manifolds,'' {\urlfont hep-th/0007213}.}

\lref\achwt{B. Acharya and E. Witten, ``Chiral Fermions from Manifolds
of $G_2$ Holonomy,'' {\urlfont hep-th/0109152}.}

\lref\ant{A. Aurilia, H. Nicolai, and P.~K. Townsend, ``Hidden
Constants: The $\Theta$ Parameter of QCD and the Cosmological Constant
of $\CN=8$ Supergravity,'' Nucl. Phys. {\bf B176} (1980) 509--522.}

\lref\atwt{M.~F. Atiyah and E. Witten, ``M-theory Dynamics on a Manifold
of $G_2$ Holonomy,'' {\urlfont hep-th/0107177}.}

\lref\bbmooy{K. Becker, M. Becker, D. Morrison, H. Ooguri, Y. Oz, and
Z. Yin, ``Supersymmetric Cycles in Exceptional
Holonomy Manifolds and Calabi-Yau 4-Folds,'' Nucl. Phys. {\bf B480}
(1996) 225--238, {\urlfont hep-th/9608116}.}

\lref\beck{K. Becker and M. Becker, ``M-theory on Eight-Manifolds,''
Nucl. Phys. {\bf B477} (1996) 155--167, {\urlfont hep-th/9605053}.}

\lref\bry{R.~L. Bryant, ``Metrics with Exceptional Holonomy,''
Ann. Math. {\bf 126} (1987) 525--576.}

\lref\chsf{B. Chibisov and M. Shifman, ``BPS-Saturated Walls in
Supersymmetric Theories'', Phys. Rev. {\bf D56} (1997) 7990--8013,
Erratum--ibid. {\bf D58} (1998) 109901, {\urlfont hep-th/9706141}.}

\lref\cjs{E. Cremmer, B. Julia, and J. Scherk, ``Supergravity Theory
in Eleven Dimensions,'' Phys. Lett. {\bf B76} (1978) 409--412.}

\lref\col{S. Coleman, ``More About the Massive Schwinger Model,''
Annals Phys. {\bf 101} (1976) 239--267.}

\lref\dalb{S.~P. de Alwis, ``A Note on Brane Tension and M-theory'',
Phy. Lett. {\bf B388} (1996) 291--295, {\urlfont hep-th/9607011}.}

\lref\dshd{B. de Wit, D.~J. Smit, and N.~D. Hari Dass, ``Residual
Supersymmetry of Compactified D=10 Supergravity,'' Nucl. Phys. {\bf
B283} (1987) 165--191.}

\lref\duffn{M.~J. Duff and P. van Nieuwenhuizen, ``Quantum
Inequivalence of Different Field Representations,'' Phys. Lett.
{\bf B94} (1980) 179--182.}

\lref\guk{S. Gukov, ``Solitons, Superpotentials, and Calibrations,''
Nucl. Phys. {\bf B574} (2000) 169--188, {\urlfont hep-th/9911011}.}

\lref\gvw{S. Gukov, C. Vafa, and E. Witten, ``CFT's From Calabi-Yau
Four-folds,'' Nucl. Phys. {\bf B584} (2000) 69--108,
Erratum--ibid. {\bf B608} (2001) 477--478, {\urlfont hep-th/9906070}.}

\lref\joyi{D.~D. Joyce, ``Compact Riemannian 7-manifolds with Holonomy
$G_2$, I.'' J. Diff. Geom. {\bf 43} (1996) 291--328.}

\lref\joyii{D.~D. Joyce, {\it Compact Manifolds with Special
Holonomy}, Oxford Univ. Press, 2000.}

\lref\hal{M. Haack and J. Louis, ``M-theory Compactified on Calabi-Yau
Fourfolds with Background Flux,'' Phys. Lett. {\bf B507} (2001)
296--304, {\urlfont hep-th/0103068}.}

\lref\hm{J.~A. Harvey and G. Moore, ``Superpotentials and Membrane
Instantons,'' {\urlfont hep-th/9907026}.}

\lref\gp{J. Gutowski and G. Papadopoulos, ``Moduli Spaces and Brane
Solitons for M-theory Compactifications on Holonomy $G_2$
Manifolds,'' Nucl. Phys. {\bf B615} (2001) 237--265,
{\urlfont hep-th/0104105}.}

\lref\page{D.~N. Page, ``Classical Stability of Round and Squashed 
Seven-Spheres in Eleven-Dimensional Supergravity,'' Phys. Rev. {\bf
D28} (1983) 2976--2982.}

\lref\pt{G. Papadopoulos and P. Townsend, ``Compactifications of D=11
Supergravity on Spaces of Exceptional Holonomy,'' Phys. Lett. {\bf
B357} (1995) 300--306, {\urlfont hep-th/9506150}.}

\lref\stong{R. Stong, ``Calculation of
$\Omega^{\hbox{\it spin}}_{11}(K(\BZ,4))$,'' in {\it Unified String
Theories}, eds. M. B. Green and D. J. Gross, World Scientific, 1986.}

\lref\vfwt{C. Vafa and E. Witten, ``A One-Loop Test of String
Duality,'' Nucl. Phys. {\bf B447} (1995) 261--270, {\urlfont
hep-th/9505053}.}

\lref\wsbg{J. Wess and J. Bagger, {\it Supersymmetry and
Supergravity, $2^{\hbox{nd}}$ ed.}, Princeton Univ. Press, 1992.}

\lref\wflux{E. Witten, ``On Flux Quantization in M-Theory and the
Effective Action,'' J. Geom. Phys. {\bf 22} (1997) 1--13,
{\urlfont hep-th/9609122}.}

\lref\wfive{E. Witten, ``Five-Brane Effective Action in M-Theory,''
J. Geom. Phys. {\bf 22} (1997) 103--133, {\urlfont hep-th/9610234}.}

\newsec{Introduction}

One route to possible $\CM$-theory phenomenology is to consider
$\CM$-theory compactifications on eleven-dimensional spaces of the
form $M_4 \times X$, where $M_4$ denotes flat Minkowski space.
When the seven-fold $X$ possesses a metric of $G_2$ holonomy,
then $M_4 \times X$ is a vacuum solution of Einstein's equation.
Further, there exists one covariantly constant spinor on $X$,
leading to an effective theory with $\CN=1$ supersymmetry in four
dimensions.  However, in contrast to $\CM$-theory
compactifications on Calabi-Yau four-folds \beck, if we
generalize this background ansatz to allow for non-zero $G$-flux
and a warped product metric on $M_4 \times X$, then no
supersymmetric vacua away from the trivial $G=0$ background exist.
This was  demonstrated, for instance, in \dshd, \achsp.

The issue of supersymmetry-breaking by $G$-flux on $M_4 \times
X$ is interesting, as this $G$-flux also generates a cosmological
constant in the four-dimensional theory \duffn, \ant.  As in the case
of compactifications on Calabi-Yau four-folds \gvw, the breaking of
supersymmetry by $G$-flux can be effectively described in the
four-dimensional theory by introducing a superpotential $W$ \guk\ for
the moduli of the compactification.

In the case of compactifications on Calabi-Yau four-folds, the
superpotentials proposed in \gvw\ have been directly verified by
a Kaluza-Klein reduction of the effective $\CM$-theory action
\hal.  One purpose of this note is to perform a similar check of
the superpotential describing $G$-flux in compactifications on
manifolds of $G_2$ holonomy.  We compare the bosonic potential
derived from $W$ with the corresponding bosonic potential
obtained from the $\CM$-theory effective action. In addition we
want to explore a phenomenon that arises in $\CM$-theory
compactification to four dimensions but not in compactification
to three dimensions on a Calabi-Yau four-fold. As in the
Freund-Rubin solution, while preserving the four-dimensional
symmetries, the $G$-field can have a component with all indices
tangent to $M_4$, possibly triggering the breaking of
supersymmetry.  We will show that the minimal superpotential that
describes the components of $G$ along $X$ actually also
incorporates the component tangent to $M_4$.  This observation
has interesting implications for the structure of the parameter
space of compactifications with $G$-flux.

The outline for this note is the following.  In Section 2 we review
the low-energy structure of Kaluza-Klein compactification of the
$\CM$-theory effective action on a smooth
manifold\foot{Compactification on a smooth $X$ is not
phenomenologically viable, since the low-energy theory
will contain only abelian vector multiplets with no charged matter.
When $X$ is allowed to have appropriate singularities, non-abelian
gauge-groups and charged chiral matter \atwt, \achwt\ can be
present, but of course the low-energy supergravity approximation is no
longer valid.} $X$ of $G_2$ holonomy.  We directly find the effective
bosonic potential for $G$.

In Section 3 we introduce and motivate the superpotential $W$.  We
then derive the bosonic potential for $G$ following from $W$.  We
find a potential which naively differs from the result of
Section 2.

Finally in Section 4, we show how the two can be reconciled.

\newsec{Kaluza-Klein Reduction of $\CM$-Theory on Manifolds of $G_2$ Holonomy}

In this section, we first review the structure of the massless $\CN=1$
multiplets that arise when eleven-dimensional supergravity is
compactified on $X$ (with $G = 0$),
as has been discussed in \pt, \hm, \gp.  The four-dimensional effective
theory possesses $b_2(X)$ abelian vector
superfields $V^j$ and $b_3(X)$ neutral chiral superfields $Z^i$.
These superfields describe massless modes of the flat $C$-field
and the metric on $X$.

To describe these modes explicitly, let us
choose bases of harmonic forms $\{ \omega_j \}$ for $\CH^2(X)$ and
$\{ \phi_i \}$ for $\CH^3(X)$.  We then make a Kaluza-Klein ansatz for $C$,
\eqn\c{
C = c^i(x) \, \phi_i + A^j_\mu(x) \, dx^\mu \^ \omega_j\,.}
This ansatz is slightly oversimplified as it applies only when the
$G$-flux is trivial.  The scalars $c^i$ and the vectors $A^j_\mu$
describe the holonomies of a flat $C$-field.  Because these holonomies
take values in $U(1)$ and not $\BR$, the fields appearing in \c\ (in
particular the scalars $c^i$) should also be regarded as taking values
in $U(1)$ rather than $\BR$.  This observation deserves emphasis---it
can also be understood by noting that under ``large''
gauge-transformations which add to $C$ a closed 3-form on $X$ of 
appropriately normalized periods, the $c^i$ undergo integral shifts.

We also note that when $X$ has $G_2$ holonomy, $b_1(X) = 0$,
for the same reasons as in the case of Calabi-Yau three-folds.  So no
harmonic 1-forms on $X$ appear in the ansatz for $C$.  Each vector
$A^j_\mu(x)$ in \c\ gives rise to one abelian vector superfield $V^j$,
and each scalar $c^i(x)$ in \c\ appears as the real component of
the complex scalar $z^i$ in the chiral superfield $Z^i$.

The corresponding imaginary components of the $z^i$ describe
massless fluctuations in the background metric on $X$.  Recall that
associated to any metric of $G_2$ holonomy on $X$ is a unique
covariantly constant (hence closed and co-closed) 3-form $\Phi$.
Given any such metric, we may associate to it the cohomology class
$[ \Phi ]$ in $H^3(X;\BR)$, and this assignment is invariant under
diffeomorphisms of $X$.  In fact, as was shown by Joyce \joyi, the
moduli space of $G_2$ holonomy metrics on $X$, modulo diffeomorphisms
isotopic to the identity, is a smooth manifold of dimension $b_3(X)$.
Further, near a point in the moduli space corresponding to the
equivalence class of metrics associated to $\Phi$, the moduli space is
locally diffeomorphic to an open ball about $[ \Phi ]$ in
$H^3(X;\BR)$.  These results imply that massless modes of the metric
on $X$ may be parametrized by introducing $b_3(X)$ scalars $s^i(x)$
defined by
\eqn\metric{
\Phi = s^i(x) \, \phi_i \,.}
(The $s^i$ are presumed to fluctuate around some point away from the
origin.)  Thus the $s^i$ in \metric\ naturally combine with the $c^i$
as $z^i = c^i + i \, s^i$.  Note that this holomorphic combination $C + i
\Phi$ on $X$ is analogous to the complexified K\"ahler class $B + i J$
familiar from compactification on Calabi-Yau three-folds.

We now recall the bosonic action of eleven-dimensional supergravity \cjs,
\eqn\S{
S_{11} = {1 \over 2 \kappa^2_{11}} \int \! d^{11} x \Big[ \sqrt{-g} R - {1
\over 2} G\^ \*  G - {1 \over 6} C \^ G \^ G \Big] \,.}
Higher derivative corrections in the $\CM$-theory effective action,
such as the $C I_8(R)$ term \vfwt, will not be relevant for the
following.  We also find it useful to introduce $T_2$ and $T_5$, the
$M2$-brane and $M5$-brane tensions, and to recall the relations
between $\kappa_{11}^2$, $T_2$, and $T_5$ (as derived, for example, in \dalb)
\eqn\ten{\eqalign{
{1 \over {2 \kappa^2_{11}}} &= {1 \over {2 \pi}} T_2 T_5 \,, \cr
T_5 &= {1 \over {2 \pi}} T^2_2 \, . \cr}}
Henceforth, we set $T_2 = 1$ to obtain the standard flux quantization
conditions on $G$.  In these units, $\kappa_{11}^2 = 2 \pi^2$ and 
$T_5 = 1 / 2 \pi$.

We wish to determine the potential induced for the moduli $Z^i$
when $G \neq 0$.  We assume that $G$ respects the Lorentz
symmetry of $M_4$ and so decomposes as \eqn\gg{ G = G_0 + G_X \,
,} where $G_0 = G|_{M_4}$ and $G_X = G|_X$.  As explained in
\wflux, Dirac quantization on $X$ generally requires that ${1 \over {2
\pi}} G_X  - {1 \over 2} \lambda$ has integral
periods, where $\lambda = p_1(X) / 2$.  So if $\lambda$ were not
even in $H^4(X;\BZ)$, our earlier assumption that $G_X = 0$ would
not have been consistent quantum mechanically.  However, the
following simple argument  (see footnote 2 of \hm) implies that
when $X$ is a spin seven-fold (one consequence of $G_2$ holonomy),
then $\lambda$ is always even. Consider $S^1 \times X$.  This is
a spin eight-fold, and $p_1(S^1 \times X) = p_1(X)$, so it
suffices to consider $\lambda$ on $S^1 \times X$. But on any spin
eight-fold, $\lambda$ being even is equivalent to the
intersection form on $H^4$ being even (for a proof of this standard
fact via index theory, see \wflux). Finally, the
intersection form on $S^1 \times X$ is even for trivial reasons.
So we learn that ${1 \over {2 \pi}} \, G_X$ must have integral periods on
$X$, consistent with $G_X = 0$.

The quadratic $GG$ term in $S_{11}$ now descends directly to a pair of
terms in the low-energy action\foot{We
will not distinguish notationally between the four-, seven-, and
eleven-dimensional Hodge $\*$, but the distinction should be clear
from context.} for the metric moduli,
\eqn\actgg{
\delta S_4^{(GG)} = - {1 \over {8 \pi^2}} \int \!\!d^4\!x 
\left[ \vol(X) \, G_0 \^ \* G_0 + \sqrt{-g_4} \, \int_X \! 
G_X \^ \* G_X \right] \, .}
A term in the low-energy action for the moduli of the $C$-field is
also induced from the $CGG$ Chern-Simons term in $S_{11}$,
\eqn\actcgg{
\delta S_4^{(CGG)} = - {1 \over {8 \pi^2}} \int \!\!d^4\!x \left[ G_0
\, \int_X \! C \^ G_X \right] \,.}
(In computing this interaction, which will be converted to an ordinary
potential in Section 4, we have dropped boundary terms from infinity
on $M_4$ which can be absorbed in the background value of $C$.  A
factor of three has arisen because the Chern-Simons term is cubic in $C$.)

Let us now adopt a slightly more suggestive notation.  We
dualize the flux $G_0$ on $M_4$ by introducing a scalar $f$ satisfying 
\eqn\gzero{
G_0 = f \, dx^0 \^ \ldots \^ dx^3\,, \; \; \* G_0 = -f \,,}
in a coframe adapted to the metric.  We also define
\eqn\th{
\theta \equiv {1 \over {4 \pi}} \int_X \! C \^ G_X \,.}
The expression $\theta$ is a seven-dimensional Chern-Simons form on
$X$.  It is not well-defined as a real number, due to the fact that
$C$ is only defined up to shifts $C \rightarrow C + \phi_i$, where the 
harmonic forms ${1 \over {2 \pi}} \phi_i$ are normalized to have
integral periods on $X$.  At first glance, one might have thought that
$\theta$ is consequently defined only modulo $\pi \cdot \hbox{integer}$.  In
fact, because the class $\lambda$ of $X$ is even, a careful treatment of
$\theta$, as given in Section 3 of \wfive, shows that
$\theta$ is actually well-defined modulo $2 \pi \cdot
\hbox{integer}$.  Hence our notation is correct in suggesting that
$\theta$ is an angle.

The four-dimensional effective potential for $C$ and $\Phi$ is then 
determined from \actgg\ and \actcgg\ after we pass to Einstein frame, 
rescaling the four-dimensional metric $g_{\mu \nu}
\rightarrow 2 \pi^2 \, \hbox{vol}(X)^{-1} \, g_{\mu \nu}$.  We find
this potential to be
\eqn\V{
V(C, \Phi) =  - {1 \over {32 \pi^6}} \, \vol(X)^3 \, f^2  + {{\pi^2}
\over 2} \, \vol(X)^{-2} \, \int_X \! G_X \^ \* G_X + {\theta \over {2
\pi}} f \,.}
The additional factors of $\vol(X)$ in the $f^2$ term
arise from the explicit factor in \actgg\ and the scaling of the
four-dimensional Hodge $\*$.  This term also has the ``wrong'' sign as 
it is really a kinetic energy, so we slightly abuse the terminology in 
referring to $V$ as a ``potential''.  The $\theta$ term, as it
descends from the eleven-dimensional Chern-Simons term, remains
independent of $\vol(X)$ under the rescaling.

\newsec{The Superpotential}

\subsec{Motivating the Superpotential}

We can now introduce the superpotential $W(Z^i)$, essentially
proposed in \guk, which describes the breaking of supersymmetry
by $G$-flux on $X$.  We consider 
\eqn\W{ W(Z^i) = {1 \over {8 \pi^2}} \int_X \left({1 \over 2} C 
+ i \, \Phi \right) \! \^ G_X \,.}
The relative factor of $1/2$ between the two terms in $W$ is required by
supersymmetry.  For under a variation \eqn\var{C \rightarrow C +
\delta C, \quad \Phi \rightarrow \Phi + \delta \Phi\,,} the superpotential
varies as \eqn\varr{W \rightarrow W + {1 \over {8 \pi^2}} \int_X
(\delta C + i\, \delta \Phi) \^ G_X\,.}  Note that a relative factor
of $2$ has appeared in the variation of the first term due to its 
quadratic dependence on $C$ and the fact that, whereas $d \Phi = 0$, 
$d C = G$.  $\delta W$ is linear in 
$\delta C + i \, \delta \Phi$ as required for holomorphy.  The condition 
for unbroken supersymmetry and zero cosmological constant of a 
four-dimensional $\CN=1$ supergravity theory with superpotential $W$ 
is that, in the vacuum, 
\eqn\brk{ W = d W = 0\,.}
The latter condition, for $W$ above, is sufficient to imply that $G_X$
must vanish in a supersymmetric vacuum with zero cosmological constant.

A discussion of the proper interpretation of the former condition is 
called for, since the term $\int_X C \^ G_X / (4 \pi)^2 =
\theta / 4 \pi$ in $W$ is only well-defined modulo 
$1 / 2 \cdot \hbox{integer}$. We have no way to pick a natural
definition of this expression as a real number, so the best we
can do is to say that all possibilities differing by $\theta
\rightarrow \theta + 2 \pi$ are allowed.  Thus, the
theory depends on an integer that is not fixed when the $C$-field
on $X$ (and its curvature $G_X$) are given.

What is the physical interpretation of this integer?
Heuristically, it corresponds to the value of the period $\int_X G_7 /
(2 \pi)^2 \; (= T_5 \int_X \! G_7 / 2 \pi \hbox{ in our conventions})$, where 
$G_7$ is the seven-form field dual to $G$.  In fact, the classical 
equation of motion $dG_7 = - {1 \over 2}  G \^
G + \dots$ (the $\dots$ being gravitational corrections that we
do not consider here) shows that $\int_X G_7 / (2 \pi)^2$ is
not constant because $G_7$ is not closed.  As observed by Page \page,
what is constant and should be quantized is rather 
$\int_X (G_7 + {1 \over 2} C \^ G) / (2 \pi)^2$, and since
${1 \over 2} \int_X C \^ G / (2 \pi)^2 \; (= \theta / 2 \pi)$ is 
anyway only defined mod an integer, we introduce no additional 
ambiguity if we take $\int_X (G_7 + {1 \over 2} C \^ G)=0$.  Thus, 
${1 \over 2} \int_X C \^ G / (2 \pi)^2$ can ``stand in'' for the 
$G_7$-flux, and the possibility of adding an integer to its value 
amounts to the possibility of shifting the $G_7$-flux by an integer 
number of quanta.

At this point, we can see that the parameter space of $M$-theory
compactifications with $G$-flux is not, as one might have
supposed, a product of the space of $C$-fields on $X$ with a copy
of $\BZ$ parametrized by the $G_7$-flux.  Rather, the parameter
space is fibered over the space of $C$-fields, with the fiber
being a copy of $\BZ$; but the fibration is non-trivial.  Consider
varying $C$ by $C\to C+C'$, where $dC'=0$.  If $C'$ has
trivial periods, this change in the $C$-field is topologically trivial.  But
${1 \over 2} \int_X C \^ G / (2 \pi)^2$  changes in a non-trivial way.  
If $G / 2 \pi$, restricted to $X$, is divisible by an integer $m$,
then the change in ${1 \over 2} \int_X C \^ G / (2 \pi)^2$ will always 
be an integer multiple of $m$, and so the change in the $G_7$-flux is 
likewise a multiple of $m$.   Thus, the only invariant under this process is
the value of $G_7$ modulo $m$.  For example, if $m=1$, $G_7$ can be
varied arbitrarily, and the overall parameter space of $C$-fields 
plus $G_7$-flux is connected.  Only when $G_X=0$ is the parameter
space what one would expect naively: a product of the space of 
$C$-fields with a copy of $\BZ$ parameterizing the $G_7$-flux.

For precise computations, it is awkward to work with $G_7$, since
(as $G$ is closed and $G_7$ is not) the theory is naturally
formulated with a three-form field $C$ and not with a dual
six-form field.  In Section 4, by treating the $C$-field quantum
mechanically, we will show how the picture we have just described
is reproduced if one works with $G$, rather than (as in the last
few paragraphs) the dual $G_7$. The need to work quantum
mechanically when one formulates the discussion in terms of $G$
should not come as a surprise; duality typically relates a
classical description in one variable to a dual quantum mechanical
description.

Previously, it was suggested \achsp\ that a superpotential
describing the effects of $G$-flux along $M_4$ and
$X$ would be
\eqn\aw{
W = \int_X G_7 + \int_X (C + i \Phi) \^ G_X\,.}
One might have naively thought that the two terms in \aw\
involving $G_7$ and $G_X$ were describing independent effects due to
$G$-flux along $M_4$ and $X$.  As we have seen, this interpretation would
be problematic because generically the parameter space of $G_7$-flux fibers
non-trivially over the space of $C$-fields on $X$.  A closely related
observation is that neither the term involving $G_7$ nor the term
involving $G_X$ individually respects holomorphy.  So the relative 
normalization of the terms is not arbitrary but fixed by holomorphy, 
contrary to what the naive interpretation would suggest.  In fact, taking 
$\int_X (G_7 + {1 \over 2} C \^ G) = 0$, we see that \aw\ corresponds, 
up to normalization, to the proposed superpotential in \W, whose
holomorphy we verified and in which $G_7$ does not explicitly appear.  

The superpotential \W\ can also be motivated and its normalization
fixed from an argument given in \gvw.  Consider an $M5$-brane
having worldvolume $\BR^{2,1} \times \Sigma$, where $\Sigma$ is a
3-cycle on $X$ which is calibrated by $\Phi$.  Such a calibrated
3-cycle is a supersymmetric 3-cycle \bbmooy\ and has minimal
volume, $\vol(\Sigma) = |\int_\Sigma \Phi|$, within its homology
class.  So the wrapped $M5$-brane appears as a BPS domain wall in
the four-dimensional theory.  The tension $\tau$ of such a BPS domain
wall in the low-energy $d=4$, $\CN=1$ Wess-Zumino model
describing the $Z^i$ is \chsf \eqn\wzt{ \tau = 2 |\Delta W|\,,}
where $\Delta W$ is the change in $W$ upon crossing the wall.

On the other hand, the $M5$-brane is a magnetic source for $G_X$, and
the class of $G_X$ must change upon crossing the wall.  This change is
$\Delta G_X = 2 \pi \, \delta_\Sigma$,
where $\delta_\Sigma$ is the fundamental class of $\Sigma$.  Hence
(assuming $C=0$) we have
\eqn\tm{\eqalign{
2 |\Delta W| &= {1 \over {(2 \pi)^2}} \left| \int_X \Phi \^ 
\Delta G_X \right| \,, \cr
&= {1 \over {2 \pi}} \left| \int_\Sigma \Phi \right| = T_5 
\vol(\Sigma) \,, \cr}}
as we expect for an M5-brane domain wall.  The dependence on $C$
follows from supersymmetry.

We now derive the the bosonic potential $U$ which follows
from the superpotential $W$.  Recall that,
in terms of the K\"ahler potential $\CK$, the metric on moduli space
$g_{i \overline{j}}$, and $W$, the potential $U$ is given
by \wsbg
\eqn\U{
U =  \exp(\CK) \Big[ g^{i \overline{j}}
D_i W \overline{D_j W} - 3 |W|^2 \Big] \,,}
where $D_i W = \partial_i W + \partial_i \CK \cdot W$ is the
covariant derivative of $W$ as a section of a line bundle over the
moduli space parametrized locally by the $Z^i$.

To evaluate $U$, we need to know $g_{i \overline{j}}$ and
$\CK$.  The metric $g_{i \overline{j}}$ can be determined most
directly from the kinetic terms of the $c^i$ fields.  These kinetic
terms arise from reducing the term $-{1 \over {8 \pi^2}} \! \int G\^
\* G$ in the action $S_{11}$.  We find these kinetic terms to be 
(in Einstein frame)
\eqn\kt{
\CL_{kin} = -{1 \over 4} \vol(X)^{-1} \,
\partial_\mu c^i \partial^\mu c^j \, \int_X \! \phi_i \^ \* \phi_j\,.}
On the other hand, the metric $g_{i \overline{j}}$ appears in the
four-dimensional action in the term $\CL_{kin} =
-g_{i \overline{j}} \partial_\mu z^i \partial^\mu \overline{z}^j$.  So
\kt\ determines the metric $g_{i \overline{j}}$ to be
\eqn\g{
g_{i \overline{j}} = \overline{\partial_j} \partial_i \CK = {1 \over
4} \vol(X)^{-1} \int_X \! \phi_i \^ \* \phi_j \,.}

We now claim that the K\"ahler potential is given, up to
shifts $\CK(Z^i,\overline{Z^i}) \rightarrow \CK + f(Z^i) +
f^*(\overline{Z^i})$, by
\eqn\k{
\CK = -3 \log{\Big[ {1 \over {2 \pi^2}} \cdot {1 \over 7} \, 
\int_X \Phi \^ \* \Phi \Big]}\,.}
The general form for $\CK$ has appeared in \hm, \gp, but we
must be careful to check the factor of $-3$
appearing in the normalization of $\CK$.  We will make this check
directly by computing $\overline{\partial_j} \partial_i \CK$.

\subsec{Mathematical Preliminaries and Computing
$\overline{\partial_j} \partial_i \CK$}

In order that the following be self-contained, we must review a few
facts about the group $G_2$ and metrics of $G_2$ holonomy (for which
a good general reference is \joyii).  As very lucidly described in
\bry, the group $G_2$ can be defined as the subgroup of $GL(7,\BR)$
preserving a particular 3-form on $\BR^7$. In terms of coordinates
$(x^1,\ldots,x^7)$ on $\BR^7$, this 3-form is
\eqn\gform{
\Phi  = \theta^{123} + \theta^{145} +
\theta^{167} + \theta^{246} - \theta^{257} - \theta^{347} -
\theta^{356} \,, }
where we abbreviate $\theta^{i_1 \ldots i_n} =
dx^{i_1} \^ \ldots \^ dx^{i_n}$.  $G_2$ also preserves the Euclidean
metric $ds^2 = (dx^1)^2 + \ldots + (dx^7)^2$ (as we expect, since
$G_2$ occurs as the holonomy of $X$ and so must be a subgroup of
$O(7)$) and hence the dual 4-form with respect to this metric,
\eqn\gpstar{
\* \Phi = \theta^{4567} + \theta^{2367} + \theta^{2345} +
\theta^{1357} - \theta^{1346} - \theta^{1256} - \theta^{1247} \,.}
We note that
\eqn\v{
dx^1 \^ \ldots \^ dx^7 = {1 \over 7} \Phi \^ \* \Phi \,,}
so that $G_2$ also preserves the orientation of $\BR^7$.

When $X$ possesses a metric of $G_2$ holonomy, then at each point $p$
of $X$ there exists a local frame in which the covariantly constant
3-form $\Phi$ takes the form in \gform\ and the metric takes the
Euclidean form.  Hence the local relation \v\ immediately implies that
\eqn\vl{
{1 \over 7} \int_X \Phi \^ \* \Phi = \vol(X)\,.}
This relation explains the factor of ${1 \over 7}$ appearing in $\CK$,
which, besides being very natural, will give the correct normalization
of $U$.

Because we are primarily interested in 3-forms on $X$, our
interest naturally lies in the $G_2$ representation $\Lambda^3 (\BR^7)^*$
consisting of rank-three anti-symmetric tensors.  $\Lambda^3
(\BR^7)^*$ decomposes under $G_2$ into irreducible representations
${\bf 1} \oplus {\bf 7} \oplus {\bf 27}$.  The trivial representation
${\bf 1}$ is generated by the invariant 3-form \gform\ which we used
to define $G_2$.  The ${\bf 7}$ arises from the fundamental
representation of $G_2$ as a subgroup of $GL(7,\BR)$ via the map
$(\BR^7)^* \hookrightarrow \Lambda^3 (\BR^7)^*$, sending
$\alpha \mapsto \* (\alpha \^ \Phi)$.  Note that $\* (\cdot \^ \Phi)$
defines a (non-zero) $G_2$-equivariant map, which must be an
isomorphism onto its image by Schur's lemma.  The ${\bf 27}$ can then
be characterized as the set of those elements $\lambda$ in 
$\Lambda^3 (\BR^7)^*$ which satisfy $\lambda \^ \Phi = \lambda \^ \*
\Phi = 0$.  This identification follows again from the fact that 
$\cdot \^ \* \Phi$ and $\cdot \^ \Phi$ are $G_2$-equivariant maps.

Now, as is familiar from the case of Calabi-Yau three-folds,
the decomposition of $\Lambda^3 (\BR^7)^*$ into
irreducible representations of $G_2$ implies a corresponding
decomposition of $\Lambda^3 T^*X$ under the holonomy.  The Laplacian
on $X$ respects this decomposition, implying a corresponding
classification of the harmonic 3-forms on $X$,
$\CH^3(X) \cong \CH^3_1(X) \oplus
\CH^3_7(X) \oplus \CH^3_{27}(X)$.  In fact, the Laplacian of $X$ as
an operator on $p$-forms depends only on the representation of the
holonomy, not on $p$.  Hence the dimension of $\CH^p_R(X)$ for
some representation $R$ depends only on $R$, not $p$.  Thus, since
$\CH^1(X) = \CH^1_7(X) = 0$, we also have that $\CH^3_7(X) = 0$.
Further, from our characterization of the ${\bf 27}$ above, we see
that $\CH^3_1(X)$ is orthogonal to $\CH^3_{27}(X)$ in the
usual inner product $(\alpha, \beta) = \int_X \alpha \^ \* \beta$.

We are now prepared to evaluate $\overline{\partial_j} \partial_i
\CK$.  First we observe that $\int_X \Phi \^ \* \Phi$ is a
 homogeneous function of the $s^i$ of degree $7 \over 3$.  This
observation follows from \vl\ and the fact that under
a scaling of the local coframe $dx^i \rightarrow \lambda dx^i$,
$\Phi$ scales as $\Phi \rightarrow \lambda^3 \Phi$, or equivalently
the $s^i$ scale as $s^i \rightarrow \lambda^3 s^i$, and $\vol(X)$ scales as
$\vol(X) \rightarrow \lambda^7 \vol(X)$.  Remembering
that $s^i = \im(z^i)$, homogeneity of $\int_X \Phi \^ \* \Phi$ in the
$s^i$ then implies that
\eqn\dK{
\partial \CK / \partial z^i = i {7 \over 2} {{\int_X \phi_i \^ \*
\Phi} \over {\int_X \Phi \^ \* \Phi}}\,.}

To evaluate a second derivative of $\CK$, we must evaluate ${\partial \over
{\partial s^i}} (\* \Phi)$ arising from the numerator of \dK.  Here
$\Phi: s^i \mapsto s^i \phi_i$ is a linear map of the (local)
coordinates $s^i$ on the moduli space, and for a fixed metric on $X$, the
Hodge $\*$ is certainly a linear operator on $\CH^3(X)$.  However,
since the operator $\*$ depends on the metric, hence the moduli $s^i$,
in general $\* \Phi(s^i)$ depends nonlinearly on the $s^i$.  Following
Joyce \joyi\ (where this derivative appears), we will denote
$\Theta(s^i) \equiv \* \Phi(s^i)$ to emphasize this nonlinearity.

Let us restrict to a local coordinate patch (diffeomorphic to $\BR^7$)
on $X$ and on this patch consider $\Theta$ as a map on a small open ball
about the canonical 3-form $\Phi$ in $\Lambda^3 T^* \BR^7$.  $\Theta$
is well-defined, since for $\Xi$ sufficiently small, a local change
of frame can always be found\foot{This statement follows from a
simple dimension count, noting that $\dim GL(7,\BR) = 49$, $\dim G_2 =
14$ and $\dim \Lambda^3 (\BR^7)^* = 35$, so that $\dim GL(7,\BR) -
\dim G_2 = \dim \Lambda^3 (\BR^7)^*$.} taking $\Phi + \Xi$ to the canonical
form \gform.  The metric associated to $\Phi + \Xi$ in the frame for which
$\Phi + \Xi$ is canonical is the Euclidean metric, and so $\Theta$
is very easy to evaluate in this frame.

We now consider the derivative $D \Theta$.  $D \Theta$ is locally
linear (i.e. linear over $C^{\infty}(\BR^7)$), so it suffices to
consider $D \Theta$ as a linear map on $\Lambda^3 (\BR^7)^*$.  Observe
that, since $\Phi$ is $G_2$-invariant, $D \Theta$ is actually
a $G_2$-equivariant map.  If we denote by $\pi^1$, $\pi^7$, and
$\pi^{27}$ the projections on $\Lambda^3 (\BR^7)^* \cong {\bf 1}
\oplus {\bf 7} \oplus {\bf 27}$, then Schur's lemma implies
that $D \Theta$ decomposes as
\eqn\dtheta{
D \Theta = D \Theta \circ \pi^1 + D \Theta \circ \pi^7 + D \Theta
\circ \pi^{27} = a \, \* \circ\pi^1 + b \, \* \circ \pi^7
+ c \, \* \circ \pi^{27} \,,}
for some constants $a$, $b$, $c$.

The above expression for $D \Theta$ certainly holds when we consider
evaluating $D \Theta$ on the restrictions of elements of $\CH^3(X)$ to
the local patch.  But the expression is also a sensible global
expression on $X$, so it must in fact hold globally.  It
remains only to evaluate the constants $a$ and $c$ (since $\CH^3_7 = 0$,
$b$ is irrelevant for us\foot{$b=1$ for the curious.}).

We determine $a$ and $c$ by explicit computation in $\Lambda^3
(\BR^7)^*$.  We fix $a$ by once more considering the scaling
$dx^i \rightarrow \lambda dx^i$, under which
$\Phi \rightarrow \lambda^3 \Phi$ and $\* \Phi \rightarrow
\lambda^4 \* \Phi$.  So $a = 4 / 3$.  To fix $c$, let us consider the
3-form $\Xi = \theta^{123} - \theta^{145}$.  Since $\Xi \^ \Phi =
\Xi \^ \* \Phi = 0$, we see that $\Xi$ transforms in the ${\bf 27}$
of $G_2$.  By a change in the frame $dx^2, dx^3 \rightarrow
(1 - \epsilon / 2) dx^2, dx^3$ and $dx^4, dx^5
\rightarrow (1 + \epsilon / 2) dx^4, dx^5$, the sum $\Phi + \epsilon \,
\Xi$ can be brought to the canonical form, to linear order in
$\epsilon$.  Dualizing and transforming back to the original frame, we
easily see that $\Theta (\Phi + \epsilon \, \Xi) = \* \Phi - \epsilon \, 
\* \Xi + \CO(\epsilon^2)$, which fixes $c = -1$.  So we finally conclude that
\eqn\derphi{
 {\partial \over {\partial s^i}} (\* \Phi) = {4 \over 3} \* \pi^1 (\phi_i)
- \* \pi^{27} (\phi_i) \,.}

This derivative in hand, we evaluate $\partial \CK / \partial
\overline{z}^j \partial z^i$ as
\eqn\ddK{\eqalign{
\overline{\partial_j}
\partial_i \CK &= - {7 \over 4} {1 \over {\int_X \! \Phi \^ \* \Phi}}
\left\{ \int_X \! \phi_i \^ \* [{4 \over 3} \pi^1(\phi_j)
- \pi^{27}(\phi_j)] - {1 \over {\int_X \! \Phi \^ \* \Phi}} \int_X \!
\phi_i \^ \* \Phi \cdot {7 \over 3} \int_X \! \phi_j \^ \* \Phi
\right\} \,, \cr
&= -{1 \over 4} \hbox{vol}(X)^{-1} \left\{ {4 \over 3} \int_X \!
\pi^1(\phi_i) \^ \* \pi^1(\phi_j) - \int_X \! \pi^{27}(\phi_i) \^ \*
\pi^{27}(\phi_j) - {7 \over 3} \int_X \! \pi^1(\phi_i) \^ \* \pi^1(\phi_j)
 \right\} \,, \cr
&= {1 \over 4} \hbox{vol}(X)^{-1} \int_X \! \phi_i \^ \* \phi_j \,, \cr}}
where we have noted that $\pi^1(\phi_i) = \Phi \cdot (\int_X \! \phi_i
 \^ \* \Phi) / (\int_X \! \Phi \^ \* \Phi)$.  Comparing to \g, we see
that $\CK$ is the properly normalized K\"ahler potential.

\subsec{Computing the Potential}

The rest of the calculation of $U$ from \U\ is now direct.
We have
\eqn\dW{
D_i W =  {1 \over {8 \pi^2}} \, \int_X \! \phi_i \^ G_X + i {7 \over
{16 \pi^2}} \left[
{{\int_X \! \phi_i \^ \* \Phi} \over {\int_X \! \Phi \^ \* \Phi}} \right]
\cdot \int_X \! ({1 \over 2} C + i \, \Phi) \^ G_X \,.}
So
\eqn\dWdW{\eqalign{
g^{i \overline{j}} D_i W \overline{D_j W} = {1 \over {7 (2 \pi)^4}} \int_X
\! \Phi \^ \* \Phi \cdot \Big\{ \int_X \! G_X \^ \* G_X &+ {21 \over 4}
\left(\int_X \! \Phi \^ \* \Phi\right)^{-1} \cdot 
\left(\int_X \! \Phi \^ G_X\right)^2 \cr
&+ \left({7 \over 2}\right)^2 \left(\int_X \! \Phi \^ \* \Phi\right)^{-1} 
\cdot \left({1 \over 2} \int_X \! C \^ G_X\right)^2 \Big\} \, .}}
Here we have noted, for instance, that $g^{i \overline{j}} \int_X
\phi_i \^ G_X \cdot \int_X \phi_j \^ G_X = {4 \over 7} \int_X \Phi \^ \*
\Phi \cdot \int_X G_X \^ \* G_X$.  This relation may be checked by expanding
$G_X = f^i(s) \, \* \phi_i$ for some functions $f^i(s)$ (although $G_X$ is
independent of the metric, the basis $\{ \* \phi_i \}$ is not).  So we find
\eqn\UU{\eqalign{
U &= {{7^3 \pi^2} \over 2} \left(\int_X \! \Phi \^ \*
\Phi \right)^{-2} \left[ {1 \over 7} \int_X \! G_X \^ \* G_X +
\left(\int_X \! \Phi \^ \* \Phi \right)^{-1} \left({1 \over 2} \int_X
\! C \^ G_X \right)^2 \right] \,, \cr
&= {{\pi^2} \over 2} \vol(X)^{-2} \left[ \int_X \! G_X \^ \* G_X +
\vol(X)^{-1} \left ({1 \over 2} \int_X \! C \^ G_X \right)^2 \right] \,. \cr}}

Comparing $V$ in \V\ to $U$ in \UU, we see that the
superpotential produces the correct $G_X^2$ term, but of course
the terms in $V$ involving $f$ do not appear in
$U$.  In terms of the angle $\theta$, we can write
\eqn\UUU{ U = {{\pi^2} \over 2} \vol(X)^{-2} \,
\int_X \! G_X \^ \* G_X + 8 \pi^6 \, \vol(X)^{-3} \, \left({\theta
\over {2 \pi}} \right)^2 \,.}

\newsec{Comparing the Potentials}

There seems to be an obvious disagreement between $V$ and $U$.  They
do not even depend on the same variables.   The component 
$G_0 = f dx^0 \^ \ldots \^ dx^3$ of the $G$-field appears in $V$ as 
\eqn\lg{ V(f) = -{1 \over {32 \pi^6}} \, \vol(X)^3 \, f^2 + 
{\theta \over {2 \pi}} \, f\,.}  Hence $V$
depends on $G_0$ as well as on the $C$-field along $X$.  By
contrast, $U$ depends only on the $C$-field along $X$.

We will now see that $V$ and $U$ can be reconciled by
treating $G_0$ quantum mechanically.  In fact, quantum
mechanically, there are only discrete allowed values for $f$;
upon setting it equal to an allowed value, $V$ coincides with
$U$.

A four-form field in 3+1 dimensions is analogous to a two-form
field -- the curvature of an abelian gauge field -- in 1+1
dimensions.  We will simply interpret \lg\ as the action for a
four-form field and treat it quantum mechanically.  (The
underlying eleven-dimensional supergravity action also contains
couplings of $G_0$ to fermions, but these are inessential for our
present purposes.) In the analogy with 1+1-dimensional abelian
gauge theory, the  term in \lg\ that is proportional to $\theta$
is analogous to the theta-angle of abelian gauge theory.

Classically, if one were allowed to just minimize $V$ with
respect to $f$, having nonzero $\theta$ would induce a non-zero
value for $f$ and hence a nonzero energy density.  This is
roughly what emerges from a proper quantum mechanical treatment
\col. If one proceeds naively,  the classical value for $f$ in
\lg\ is
\eqn\gflux{ 
f = 8 \pi^5 \, \vol(X)^{-3} \, \theta \,,} 
leading to the induced energy density
\eqn\vac{ E(\theta) = 8 \pi^6 \, \vol(X)^{-3} \, \left( {\theta \over
{2 \pi}} \right)^2 .}
Quantum mechanically, one really wants the energies of all of the states;
one finds \col\ that the quantum states of this system are
labeled by an integer $n$, with the energy density of the
$n^{th}$ state being \eqn\nvac{ E_n(\theta) = 8 \pi^6 \, \vol(X)^{-3} \,
\left(n + {\theta \over {2 \pi}} \right)^2\, .}

A different way to describe the above results is that \vac\ is,
indeed, the correct formula for the energy, but in interpreting
$\theta/2\pi$ as a real number, one must include all
possibilities, differing by the possible addition of an integer.
Of course, one could, for any given $\theta$ pick the --
generically unique -- integer that minimizes the energy. However,
we prefer to show that our two potentials $V$ and $U$
agree (when $V$ is treated quantum mechanically) without imposing
any such restriction.  For this, we simply use \vac, but
accepting all real lifts of $\theta$ differing by multiples of $2
\pi$,  as we have anyway done throughout this paper.

Comparing to \UUU, we see that the vacuum energy density
$E(\theta)$ due to $G_0$ is exactly the second term in
$U$, so that the superpotential $W$ indeed captures
the effects of supersymmetry-breaking by both $G_X$ and $G_0$.

Finally, we note that the potential $U$ is positive-definite.  In the 
regime of large $\vol(X)$, for which our
effective $\CM$-theory action is valid, no vacuum exists and we see a
runaway to infinite volume, a familiar situation in supergravity
compactifications.  We might also consider contributions to the potential
from membrane instantons wrapping calibrated 3-cycles $\Sigma$ on $X$,
but the leading instanton contribution to the potential \hm\ in the
large volume regime is of order $e^{-\vol(\Sigma)}$ and so will not help
to stabilize the runaway.

\bigbreak\bigskip\bigskip\centerline{{\bf Acknowledgements}}\nobreak
The work of C.B. is supported by an NSF Graduate Fellowship and 
under NSF Grant PHY-9802484.  The work of E.W. is supported in part by
NSF Grant PHY-0070928.

\listrefs

\end